\documentclass{crm-article}
\usepackage{float}
\usepackage{hyperref}
\usepackage{tabularx}
\usepackage{url}
\usepackage{verbatim}
\usepackage{graphicx}
\usepackage{cite}

\begin{document}








\title{Оптимизированные методы машинного обучения для исследования термодинамического поведения сложных спиновых систем}
\titleeng{Optimized Machine Learning Methods for Studying the Thermodynamic Behavior of Complex Spin Systems}

\thanks{Исследование выполнено за счет гранта Российского научного фонда №~25-21-00286, \url{https://rscf.ru/project/25-21-00286/}.}
\thankseng{The study was funded by the Russian Science Foundation, grant No.~25-21-00286, \url{https://rscf.ru/project/25-21-00286/}.}

\author[1]{\firstname{Д.\,Ю.}~\surname{Капитан}}
\authorfull{Д.\,Ю. Капитан}
\authoreng[1]{\firstname{D.}~\surname{Kapitan}}
\authorfulleng{Dmitrii Kapitan}
\email{kapitandmitrii@gmail.com}

\author[1,2]{\firstname{П.\,А.}~\surname{Овчинников}}
\authorfull{П.\,А. Овчинников}
\authoreng[1,2]{\firstname{P.}~\surname{Ovchinnikov}}
\authorfulleng{Pavel Ovchinnikov}


\author[1,2]{\firstname{К.\,С.}~\surname{Солдатов}}
\authorfull{К.\,С. Солдатов}
\authoreng[1,2]{\firstname{K.}~\surname{Soldatov}}
\authorfulleng{Konstantin Soldatov}


\author[3]{\firstname{П.\,Д.}~\surname{Андрющенко}}
\authorfull{П.\,Д. Андрющенко}
\authoreng[3]{\firstname{P.}~\surname{Andriushchenko}}
\authorfulleng{Petr Andriushchenko}

\author[4]{\firstname{В.\,Ю.}~\surname{Капитан}}
\authorfull{В.\,Ю. Капитан}
\authoreng[4]{\firstname{V.}~\surname{Kapitan}}
\authorfulleng{Vitalii Kapitan}
\email{kapitanvy@gmail.com}


\affiliation[1]{Институт наукоемких технологий и передовых материалов, 
Дальневосточный федеральный университет,\protect\\
о.~Русский, пос.~Аякс, д.~10, Владивосток, 690922, Россия}
\affiliationeng[1]{Institute of High Technologies and Advanced Materials, 
Far Eastern Federal University,\protect\\
Russky Island, Ajax~10, Vladivostok 690922, Russia}

\affiliation[2]{Институт прикладной математики, 
Дальневосточное отделение Российской академии наук,\protect\\
ул.~Радио, д.~7, Владивосток, 690041, Россия}
\affiliationeng[2]{Institute of Applied Mathematics, 
Far Eastern Branch, Russian Academy of Sciences,\protect\\
Radio~7, Vladivostok 690041, Russia}

\affiliation[3]{Heidelberg Institute of Global Health, Heidelberg University Hospital, Heidelberg University,\protect\\
 INF 130.3, Heidelberg 69120, Germany}
\affiliationeng[3]{Heidelberg Institute of Global Health, 
 Heidelberg University Hospital, Heidelberg University,\protect\\
 INF 130.3,  Heidelberg, 69120, Germany}

\affiliation[4]{Institute for Functional Intelligent Materials, 
National University of Singapore,\protect\\
21 Lower Kent Ridge Road, 119077, Singapore}
\affiliationeng[4]{Institute for Functional Intelligent Materials, 
National University of Singapore,\protect\\
21 Lower Kent Ridge Road, Singapore 119077, Singapore}

\begin{abstract}

В настоящей работе проводится систематическое исследование применения сверточных нейронных сетей (CNN) в качестве эффективного и универсального инструмента для анализа критических и низкотемпературных фазовых состояний в моделях спиновых систем.
Рассматривается задача расчета зависимости средней энергии $\langle E\rangle_T$ от пространственного распределения обменных интегралов ${J_k}$ для модели Эдвардса–Андерсона на квадратной решётке с фрустрированными взаимодействиями.
Реализуется единый свёрточный классификатор фазовых состояний ферромагнитной модели Изинга на квадратной, треугольной, гексагональной и кагоме решётках, обученный на конфигурациях, сгенерированных кластерным алгоритмом Свендсена–Ванга.
Температурные профили усреднённой апостериорной вероятности
высокотемпературной фазы, вычисленные этим классификатором, образуют
чёткие S-образные кривые с пересечением вблизи теоретических критических
температур и позволяют установить значение $T_c$ для решётки кагоме без
дополнительного дообучения. Показано, что свёрточные модели позволяют существенно снизить среднеквадратичную ошибку (RMSE) по сравнению с полносвязными архитектурами и эффективно улавливают  сложные связи между термодинамическими характеристиками и структурой магнитных коррелированных систем.
\end{abstract}





\begin{abstracteng}
This paper presents a systematic study of the application of convolutional neural networks (CNNs) as an efficient and versatile tool for the analysis of critical and low-temperature phase states in spin system models. The problem of calculating the dependence of the average energy $\langle E\rangle_T$ on the spatial distribution of exchange integrals ${J_k}$ for the Edwards–Anderson model on a square lattice with frustrated interactions is considered. 

We further construct a single convolutional classifier of phase states of the ferromagnetic Ising model on square, triangular, honeycomb, and kagome lattices, trained on configurations generated by the Swendsen--Wang cluster algorithm. Сomputed temperature profiles of the averaged posterior probability of the high-temperature phase, form clear S-shaped curves that intersect in the vicinity of the theoretical critical temperatures and allow one to determine $T_c$ for the kagome lattice without additional retraining. 

It is shown that convolutional models substantially reduce the root-mean-square error (RMSE) compared with fully connected architectures and efficiently capture complex correlations between thermodynamic characteristics and the structure of magnetic correlated systems.
\end{abstracteng}

\keywordeng{Ising model}

\keywordeng{Spin glass}

\keywordeng{Machine Learning}

\keywordeng{Convolutional Neural Networks}

\maketitle


\paragraph{Введение}

Описание природы низкотемпературной фазы в спиновых стёклах и фрустрированных спиновых системах остаётся одной из ключевых задач статистической физики \cite{edwards1975theory,parisi1983order,jesi2016,diep2025}. Изучение термодинамических характеристик моделей с взаимодействием между ближайшими соседями, как правило, опирается на анализ статистической суммы, содержащей информацию обо всём множестве допустимых микросостояний системы \cite{fisher1988,yucesoy2012}. Однако, с увеличением размера системы число состояний растёт экспоненциально, что делает прямое вычисление статистической суммы для общего случая практически недостижимым \cite{lukas2014}. Дополнительные трудности связаны с характерными для фрустрированных систем большими временами релаксации, сложным энергетическим ландшафтом и макроскопической вырожденностью основного состояния \cite{krzakala2000,arguin2010uniqueness,arguin2019relation,itoi2021}.   Хотя достижение равновесия требует учёта большого числа микросостояний, рост вычислительных мощностей и совершенствование вычислительных методов позволили частично компенсировать возрастающую сложность задач, связанных с моделированием сложных спиновых систем \cite{nefedev2013concentration, vasil2020numerical, makarova2023canonical, lgotina2025non, ovchinnikov2025hierarchy}.

Из-за таких ограничений, важное значение приобретают подходы, позволяющие исследовать фазовые переходы и критические явления без явного вычисления статистической суммы. Особый интерес при этом представляют  алгоритмы распознавания фазовых состояний, которые опираются на статистику конфигураций и не требуют отдельной перенастройки под каждую конкретную геометрию решётки, размер системы или температурный диапазон.  
Методы машинного обучения предлагают инновационные подходы к решению подобных проблем, позволяя выявлять скрытые закономерности и предсказывать новые физические явления \cite{carrasquilla2017machine, Dean2018,mcnaughton2020,Bukov2021,kapitan2021numerical,zhang2021,Korol2022,fan2023}. 
В частности, сверточные нейронные сети (CNN) особенно хорошо подходят для обработки пространственно структурированных данных. Их архитектура, включающая последовательное применение свёрточных слоёв, позволяет автоматически извлекать иерархии признаков \cite{Kapitan2023,jiang2024point}.

В настоящей работе проводится систематическое исследование применения сверточных нейронных сетей в качестве эффективного и универсального инструмента для анализа критических и низкотемпературных фазовых состояний в моделях спиновых систем.
Были рассмотрены две взаимосвязанные задачи. Во-первых, восстановление функциональной зависимости между средней энергией системы $\langle E\rangle_{T}$ и пространственным распределением обменных интегралов \(J_k = f_J(x_k, y_k)\) на квадратной решётке спинового стекла. A также распознавание фазовых состояний в ферромагнитных решётках различных геометрий, что естественным образом позволяет определять значение критических температур без явного вычисления корреляционных функций или гамильтониана. Анализ качества предсказаний в зависимости от размера системы, типа распределения \(J_k\) и вида решётки показывает, что один и тот же тип архитектуры сверточных нейронных сетей (CNN) способен улавливать сложные связи между термодинамическими характеристиками и структурой магнитных коррелированных систем. Такой подход допускает обобщение на более широкий класс задач, включая модели с иными типами взаимодействий, фрустрациями, размерностями, а также для изучения других термодинамических характеристик \cite{Butler2018, shiina2024super,fu2024,ju2025}.

\paragraph{Модель и обучающие данные}

В настоящем исследовании была рассмотрена двумерная спиновая модель Изинга, определяемая Гамильтонианом вида: $\mathcal{H} = - \sum_{{}} J_{ij} S_i S_j$, где $S_i = \pm 1$ - спины, расположенные на узлах различных типов решёток и взаимодействующие по заданным обменным интегралам  $J_{ij}$ между ближайшими соседями с учетом периодических граничных условий. Константы $J_{ij}$ определяют характер взаимодействия: случайное распределение знаков $J_{ij} = \pm 1$ соответствует фрустрированным спин-стекольным системам, а фиксированные значения $J_{ij} = +1$ — регулярным ферромагнитным решёткам. Таким образом, выбранная физическая модель непосредственно формирует архитектуру и формат обучающих данных: в первом случае обучающая выборка состоит из пространственно-случайных конфигураций взаимодействий, во втором — из конфигураций спинов при разных температурах для регулярных структур. Подход позволяет единообразно рассматривать широкий спектр физических задач, от анализа фазовых переходов до изучения корреляций в системах с различной топологией и типом фрустрации.

\subparagraph{Регрессия энергии спинового стекла}

В этой задаче исследуется модель Эдвардса–Андерсона на квадратной решётке с фрустрированными, случайно выбранными $J_{ij} = \pm 1$ \cite{edwards1975theory}. Для каждой системы с заданным распределением $J_{ij}$ и температуры $T$ вычисляется средняя энергия
\begin{gather} \label{E_T_sg}
\langle E \rangle_T = \frac{1}{N}\langle \mathcal{H} \rangle_T,
\end{gather}
которая и используется в качестве целевой переменной для задачи регрессии. 

Для установления зависимости между пространственным распределением обменных взаимодействий и термодинамическими характеристиками было реализовано и обучено несколько архитектур сверточных нейронных сетей. Применение методов глубокого обучения в задачах регрессии позволяет выявлять скрытые структурные закономерности и анализировать их вклад в формирование макроскопических свойств спиновых систем.

Для обучения моделей были подготовлены отдельные выборки для обучения, валидации и тестирования. Рассматривались две модели спинового стекла размерами $6\times6$ и $10\times10$, для которых генерировались различные случайные распределения интегралов обменного взаимодействия $J_{ij}$. Для каждой конфигурации вычислялись значения физических характеристик при $60$ температурах в диапазоне от $0.1$ до $6$ с шагом $0.1$. В итоге, на вход нейронной сети подавались данные $\{J_{ij}, T\}$, а на выходе предсказывались значения средней энергии от температуры $\langle E \rangle_T$.

В архитектуре сверточных нейронных сетей обменные взаимодействия были разложены на два канала: $J_{hor}$, описывающий горизонтальные связи, и $J_{ver}$, описывающий вертикальные взаимодействия, что позволило явно учитывать пространственную структуру решётки.

Точным методом трансфер-матриц \cite{padalko2022parallel} были рассчитаны значения средней энергии от температуры для $6860$ конфигураций спинового стекла размера $10\times10$. Так как значения средней энергии каждой системы были рассчитаны для $60$ значений температур, это сформировало обучающий массив размером $411600$ строк. Аналогично, для решётки $6\times6$ было сформировано $2010$ независимых конфигураций и датасет размером $120600$ соответственно.

\subparagraph{Классификация фазовых состояний}

Для задачи классификации фазовых состояний использовался вышеописанный  гамильтониан, где все взаимодействия были ферромагнитными $J_{ij} = 1$ для всех рёбер. Анализ выполнялся на нескольких типах пространственных решёток: квадратной, треугольной, гексагональной и кагоме. При этом конфигурации на решётке кагоме использовались исключительно для независимого тестирования и не входили в обучающий набор, что позволило оценить обобщающую способность модели.

Для обучения сверточной нейронной сети был сформирован набор данных, включающий конфигурации моделей Изинга на квадратных ($32\times32$, $48\times48$, $56\times56$), треугольных ($32\times32$, $48\times48$, $56\times56$) и гексагональных ($22\times22$, $34\times34$, $42\times42$) решётках, сгенерированных кластерным алгоритмом Свендсена-Ванга~\cite{SwendsenWang1987}. Для каждого типа решётки рассматривались два непересекающихся температурных диапазона -- ниже и выше критической температуры $T_c$. Конфигурации из узкой окрестности $T_c$ сознательно исключались, чтобы уменьшить риск переобучения, вызванного высокой вариативностью состояний в критическом режиме.

Для каждого фиксированного значения температуры и размера решётки отбиралось не более $600$ независимых конфигураций, что обеспечивало баланс между температурными классами. В итоге было использовано $43$ температурных значения для квадратной, $47$ -- для треугольной и $58$ -- для гексагональной решётки. Совокупный набор данных включал $444$ комбинации «температура × размер решётки» и $266400$ конфигураций.

\paragraph{Исследование термодинамического поведения спиновых стекол}

Для решения задачи регрессии средней энергии от температуры $\langle E\rangle$ см. формулу \eqref{E_T_sg}, спиновое стекло рассматривалось как взвешенный граф, в котором значения обменного взаимодействия представлены значениями ребер, а архитектура графа соответствует решетке. Для двумерной системы размером $N=L\times L$ возможно $2^{2N}$ различных распределений значений $J_{ij}$. Эти конфигурации охватывают весь спектр — от полностью антиферромагнитного состояния, при котором каждое значение $J_k=-1$, что приводит к $\sum_{k=1}^{2N}J_k=-2N$, до полностью ферромагнитного случая, где $J_k=1$ для всех $k$, и, соответственно, $\sum_{k=1}^{2N}J_k=2N$. 

С помощью сверточной нейроной сети мы хотим найти функциональную зависимость между средней энергией $\langle E\rangle_{T}$ и пространственным распределением обменных  интегралов  на квадратной решетке спинового стекла $J_k=f_{J}(x_k,y_k)$. Здесь $x_k$ и $y_k$ представляют координаты связи $k$, $J_k$ — значение связи, а $f_{J}$ — функция пространственного распределения значений связей спинового стекла. Для решения этой задачи было предложено использовать  несколько архитектур CNN, описание которых приведено ниже.

\begin{figure}[t]

\centering
\includegraphics[width=0.55\textwidth]{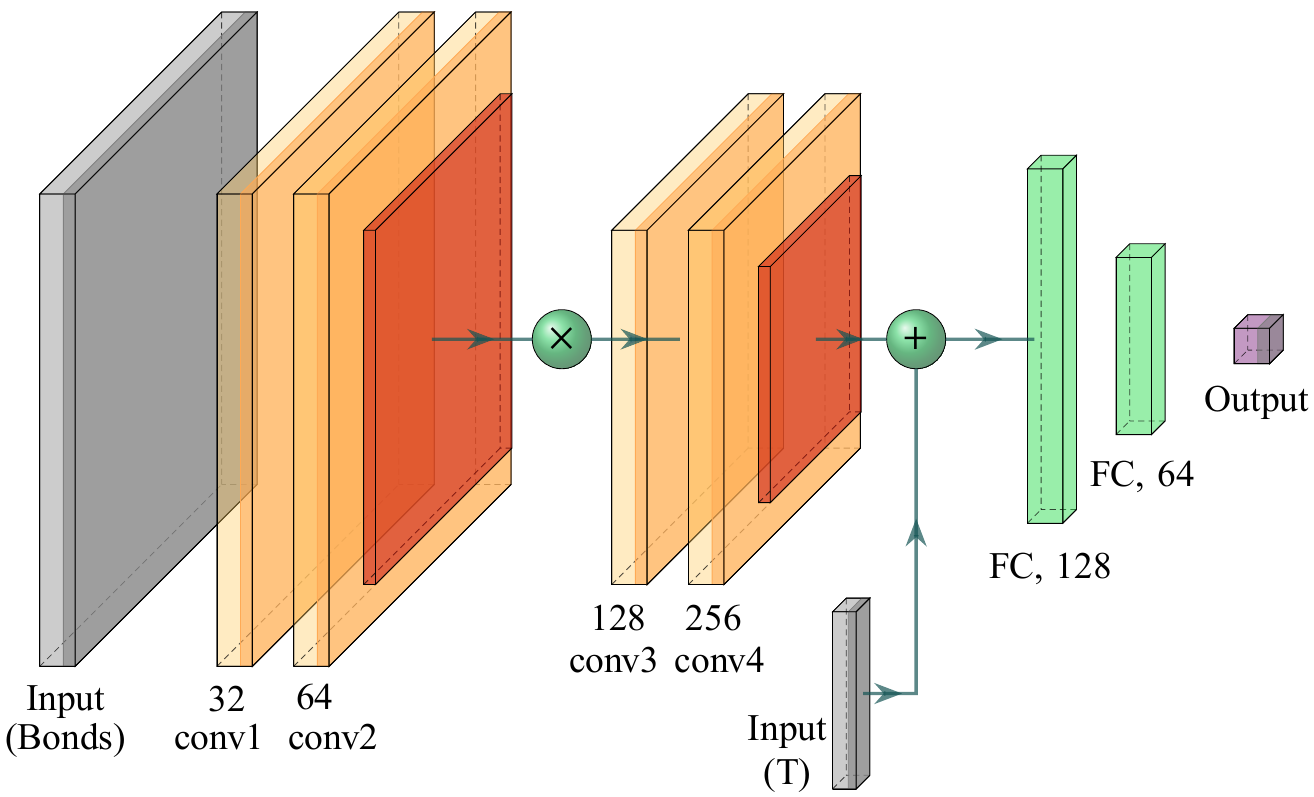}
\smallskip
\caption{Предложенная архитектура CNN1.}
\label{img:cnn1_architecture}
\end{figure}

Архитектура первой предложенной нейросетевой модели изображена на рис. \ref{img:cnn1_architecture}. Она построена таким образом, чтобы обрабатывать конфигурации связей посредством последовательности операций свертки и пулинга (pooling). Изначально сеть использует два свёрточных слоя ($Conv2D$) для извлечения низкоуровневых признаков из входных данных о связях, за которыми следуют слои пулинга с функцией максимума ($MaxPool$), которые уменьшают пространственную размерность, сохраняя при этом основную информацию. Для улучшения обобщения и предотвращения переобучения на этом этапе используется слой Dropout. Дальнейшие свёрточные слои и слои пулинга постепенно уточняют представления признаков и сжимают данные. Слой глобального усреднения ($GAP$) затем агрегирует эти характеристики в компактное векторное представление. Затем этот вектор характеристик соединяется с вспомогательным входом ($Concatenate$), соответствующим температуре, что позволяет сети интегрировать как структурную, так и термодинамическую информацию. Полученный в результате комбинированный вектор проходит через полносвязные слои ($Dense$) для выполнения регрессии, что дает прогнозы средних уровней энергии. Такая архитектура облегчает эффективное моделирование поведения энергии в системах спинового стекла за счет совместного использования данных о конфигурациях связей и температуре.

Детально, архитектура, представленная на рис. \ref{img:cnn1_architecture} описывается следующим образом: $Conv2D: (32, 3 \times 3)$ $\rightarrow$ $Conv2D: (64, 3 \times 3)$ $\rightarrow$ $MaxPool: (2 \times 2)$ $\rightarrow$ $Dropout: (0.2)$ $\rightarrow$ $Conv2D: (128, 3 \times 3)$ $\rightarrow$ $Conv2D: (256, 3 \times 3)$ $\rightarrow$ $MaxPool: (2 \times 2)$ $\rightarrow$ $GAP$ $\rightarrow$ $Concatenate: (\text{Temp input})$ $\rightarrow$ $Dense: (128)$ $\rightarrow$ $Dense: (64)$ $\rightarrow$ $Dense: (1)$.

\begin{figure}[!h]

\centering
\includegraphics[width=0.70\textwidth]{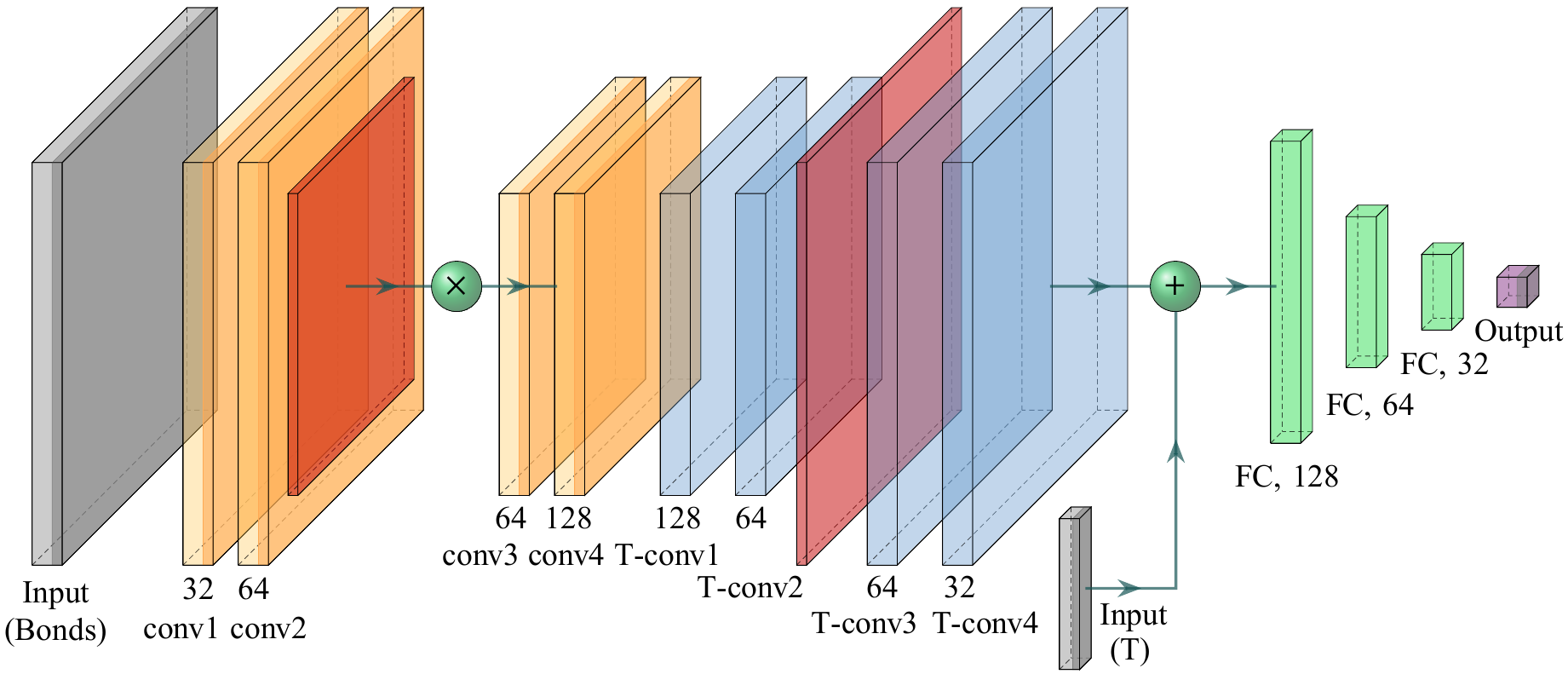}
\smallskip
\caption{Предложенная архитектура CNN2.}
\label{img:cnn2_architecture}
\end{figure}

Вторая модель CNN, представленная на рис. \ref{img:cnn2_architecture}, объединяет сверточные, повышающей дискретизации (upsampling) и полносвязные слои для эффективного захвата и обработки сложных паттернов, присутствующих в конфигурациях связей. Отличительной особенностью этой архитектуры является включение слоев повышающей дискретизации, реализованных с помощью транспонированных сверточных операций (Conv2DTranspose), которые служат для реконструкции пространственных измерений карт признаков. Этот механизм позволяет модели восстанавливать представления с высоким разрешением и сохранять мелкие детали из входных данных. После этапов повышающей дискретизации, модель применяет глобальное усреднение для сжатия пространственно расширенных карт признаков в компактное векторное представление. Затем этот вектор признаков соединяется с вспомогательным скалярным входом, соответствующим температуре. Впоследствии объединенное представление обрабатывается через серию полносвязанных слоев для выполнения задачи регрессии.

Эту архитектуру, показанную на рис. \ref{img:cnn2_architecture} можно описать следующим образом: $Conv2D: (32, 3 \times 3)$ $\rightarrow$ $Conv2D: (64, 3 \times 3)$ $\rightarrow$ $MaxPool: (2 \times 2)$ $\rightarrow$ $Dropout: (0.2)$ $\rightarrow$ $Conv2D: (64, 3 \times 3)$ $\rightarrow$ $Conv2D: (128, 3 \times 3)$ $\rightarrow$ $Conv2DTranspose: (128, 3 \times 3)$ $\rightarrow$ $Conv2DTranspose: (64, 3 \times 3)$$\rightarrow$ $UpSampling: (2 \times 2)$ $\rightarrow$ $Conv2DTranspose: (64, 3 \times 3)$ $\rightarrow$ $Conv2DTranspose: (32, 3 \times 3)$ $\rightarrow$ $GAP$ $\rightarrow$ $Dense: (128)$ $\rightarrow$ $Dense: (64)$ $\rightarrow$ $Dense: (32)$.

\begin{figure}[H]
    \centering
    \subfloat[]{%
        \includegraphics[width=0.40\textwidth]{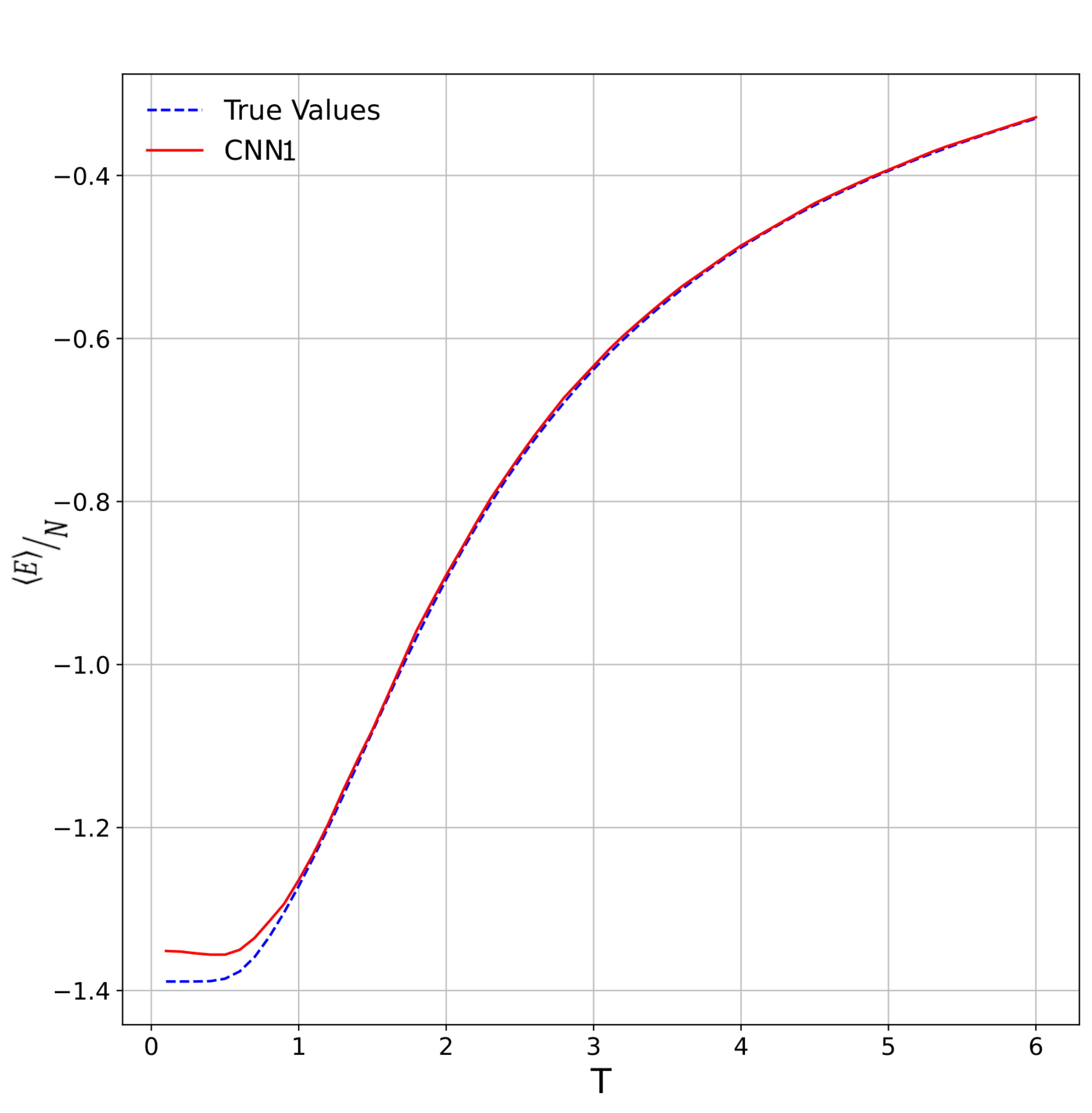}%
    }\hfill
    \subfloat[]{%
        \includegraphics[width=0.40\textwidth]{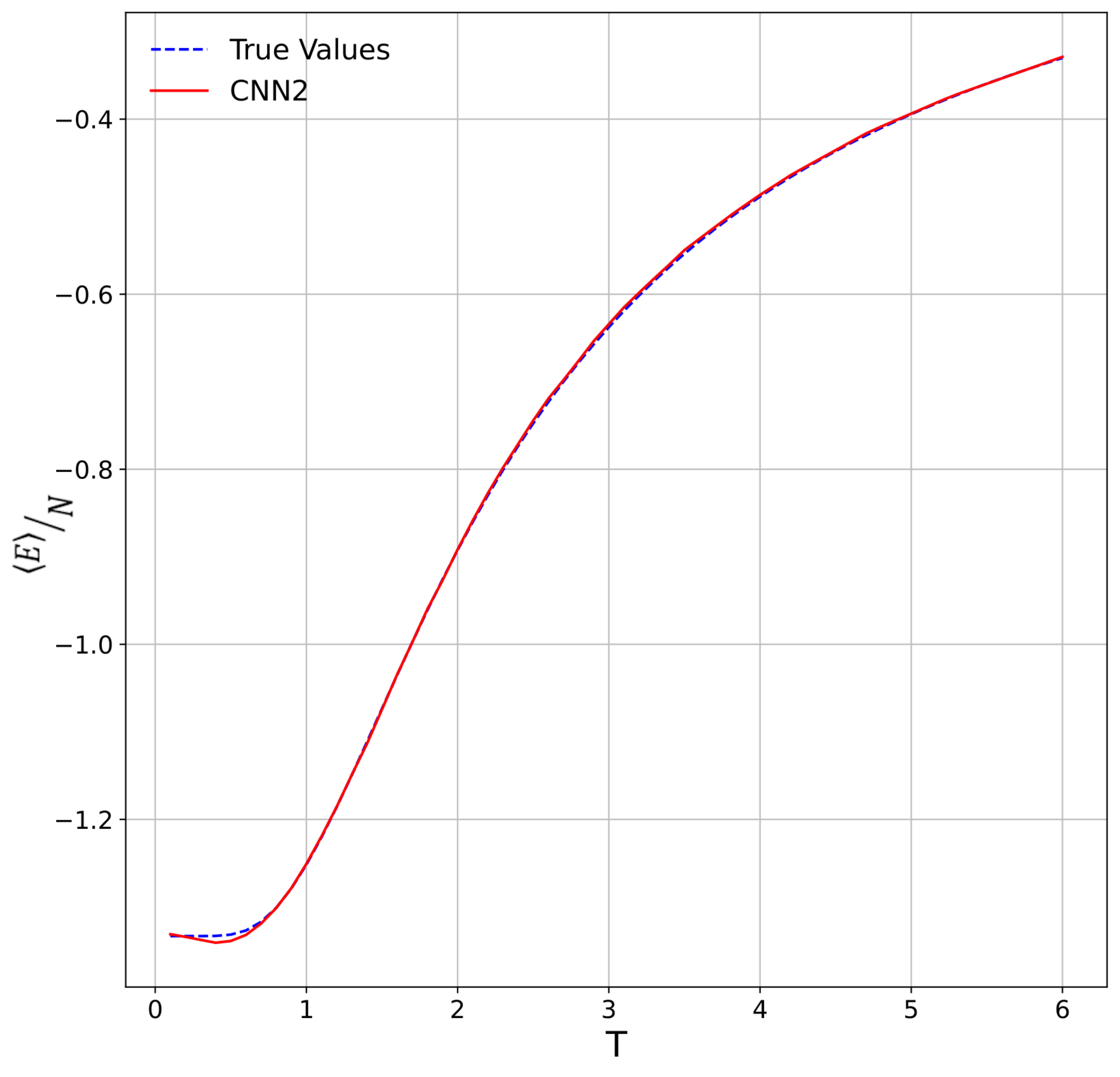}%
    }\\
    \subfloat[]{%
        \includegraphics[width=0.40\textwidth]{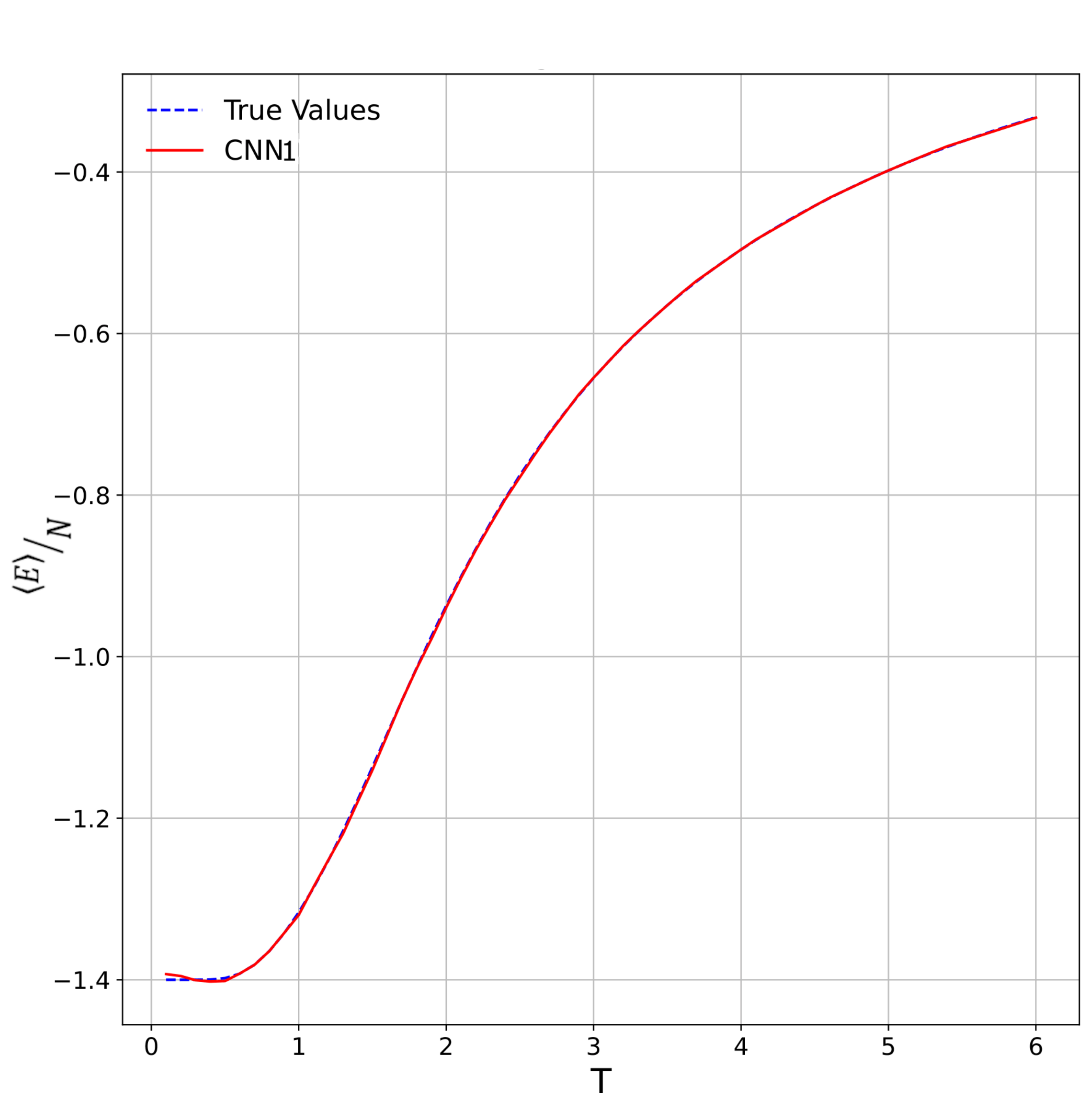}%
    }\hfill
    \subfloat[]{%
        \includegraphics[width=0.40\textwidth]{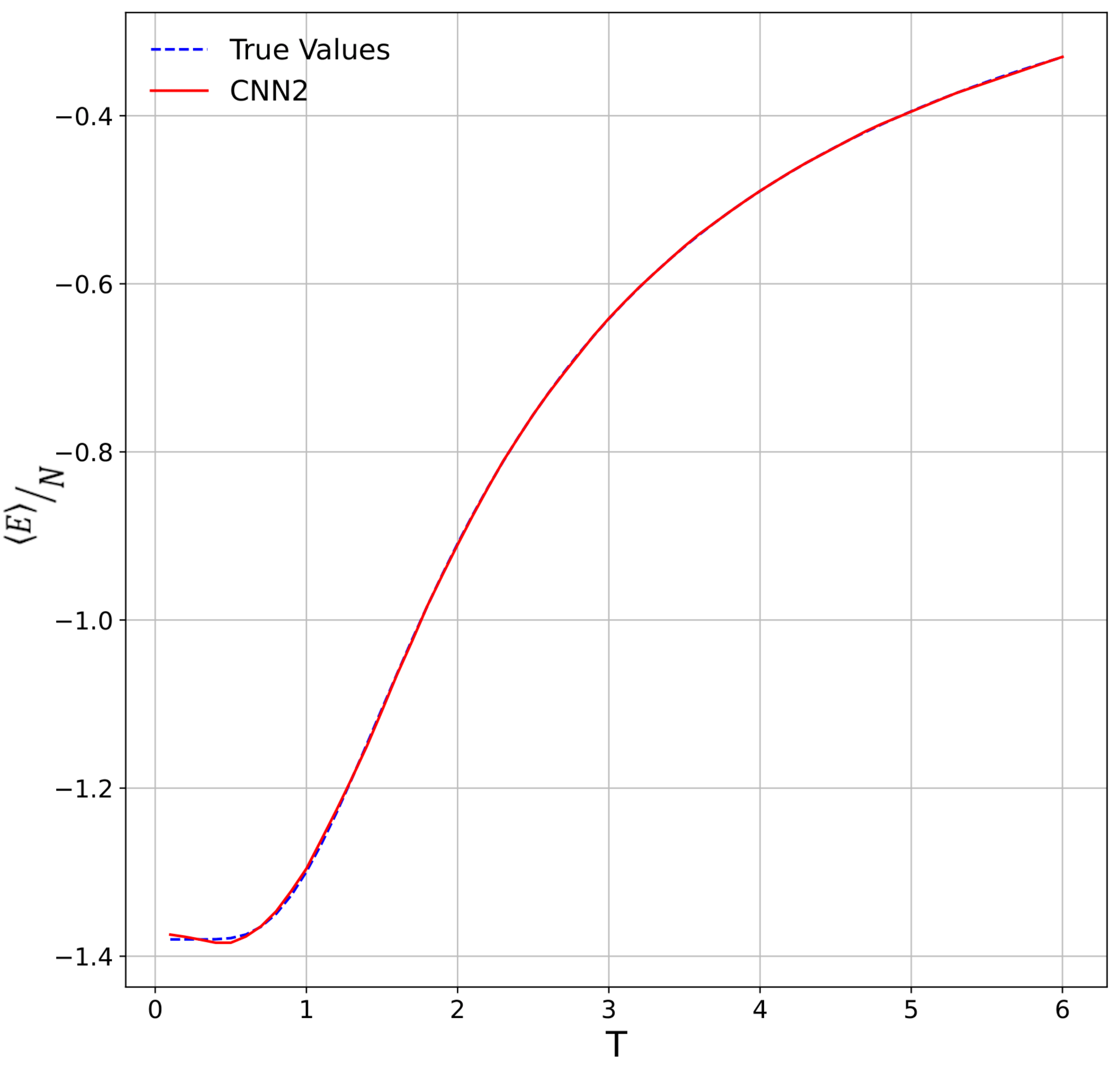}%
    }

    \caption{\centering Сравнение зависимостей средних значений энергии от температуры
    для случайных конфигураций, полученных с помощью метода трансфер матриц
    (True Values) и нейронных сетей (CNN1 и CNN2) для систем разного размера:
    a) $6\times6$, CNN1; b) $6\times6$, CNN2; c) $10\times10$, CNN1; d) $10\times10$, CNN2.}
    \label{img:cnn_predictions}
\end{figure}

Для оценки точности работы CNN были также обучены полносвязные нейронные сети с разным количеством и размерами скрытых слоев — FC1 и FC2, а также архитектуры глубоких нейронных сетей с двумя уровнями абстракции спиновой решетки — CC1 и CC2 (от англ. Custom Connected Neural Networks, CC) из \cite{andriushchenko2022new}.  Кроме того, для оптимизации CNN был апробирован подход с масштабированием градиента, т.н. метод модификации ландшафта (LM) \cite{choi2024improved,kapitan2025adaptive}. Метод LM улучшает оптимизацию путем преобразования целевой функции $g(x)$ в модифицированную форму $\hat{g}(x)$, управляемую параметрами $a$ и пороговым значением $c$. Это может помочь оптимизаторам, например, Adam лучше избегать локальных минимумов и седловых точек, что приводит к более быстрой сходимости к глобальному или близкому к нему локальному оптимуму. В нашем случае процедура масштабирования выглядит следующим образом: градиент $\nabla g_t$ масштабируется с помощью функции преобразования $f$  текущей потери $rl_{t}$ и \mbox{параметра $c_t$ на шаге $t$}:
 \begin{gather}\label{lm}
       {\nabla \hat{g}_t} = \frac{\nabla g_t}{a f((rl_{t} - c_t)_+) + 1},
   \end{gather}
где $c_t$ было установлено близко к минимальному значению функции потерь.

В данном исследовании мы использовали соотношение данных $0,8:0,15:0,05$ для обучения, валидации и тестирования соответственно. Модели сначала обучались на наборе данных для размеров $6\times6$, а затем дополнительно обучались на наборе данных для размеров $10\times10$. Количество эпох для каждого этапа обучения составляло $10$, а размер батча — $256$.
Среднеквадратичные ошибки (RMSE) средней энергии были рассчитаны для выходных значений различных CNN от эталонных значений, полученных с помощью метода трансфер-матриц. RMSE рассчитывалась для $6\times6$ и $10\times10$ по всему тестовому набору данных и усреднялась по температуре и различным тестовым конфигурациям распределения интеграла обмена $J$. Полученные значения RMSE в зависимости от архитектуры CNN и размера системы представлены в таблице \ref{tab:tab2}. 

Из Таблицы \ref{tab:tab2} можно сделать вывод что сверточные нейронные сети (CNN1, CNN2) и их модификации с масштабированием градиента (CNN1LM, CNN2LM) значительно превосходят полносвязные и кастомные модели, снижая ошибку предсказания на порядок (например $0.0017$ у CNN2LM против $0.0382$ у полносвязной сети для $10\times10$). Такой прирост видимо связан с эффективным извлечением пространственных корреляций и адаптацией к структуре взаимодействий. В свою очередь, добавление масштабирования градиента обеспечивает дополнительную устойчивость при росте размера системы и стабильное снижение RMSE.

\graphicspath{{pdf/}} 
\begin{table}[H]
\centering
\caption{Сравнение точности предсказаний средней энергии на основе RMSE для полносвязных сетей (FC1, FC2), сетей с кастомной архитектурой (CC1, CC2) , сверточных сетей (CNN1, CNN2) и сверточных сетей с масштабированием градиента (CNN1LM, CNN2LM) .}
\medskip
\begin{tabular}{|c|c|c|c|c|c|c|c|c|}
\hline
N & FC1 & FC2 & CC1 & CC2 & CNN1 & CNN2 & CNN1LM & CNN2LM \\
\hline
$6\times6$ & 0.0555 & 0.0557 & 0.0413 & 0.0491 & 0.0118 & \bf{0.0025} & 0.0034 & 0.0044\\
\hline
$10\times10$ &  0.0377 & 0.0382 & 0.0261 & 0.0302 & 0.0018 & 0.0019 & 0.0031  & \bf{0.0017}\\
\hline
\end{tabular}
\label{tab:tab2} 
\end{table}

На рис. \ref{img:cnn_predictions} показаны сравнения вычисления средней энергии с использованием точного метода трансфер матриц и CNN различных архитектур (CNN1, CNN2) для моделей из тестового датасета.
Анализ графиков (рис.~\ref{img:cnn_predictions}) показывает, что все архитектуры сверточных сетей обеспечивают высокую точность расчета средней энергии при $T>1$, причем различия между архитектурами начинают проявляться только в области низких температур ($T<1$) наиболее сложной для моделирования. Модель CNN2 демонстрирует наилучшее совпадение с вычисленными значениями, что подтверждает важность оптимального выбора архитектуры сверточных сетей.

\paragraph{Единый сверточный классификатор фаз на различных решётках Изинга}

Рассмотрим задачу распознавания фазовых состояний в двумерных моделях Изинга на решётках различных геометрий. Целью является построение единого классификатора, который по конфигурации спинов определяет принадлежность системы к низкотемпературной или высокотемпературной фазе и тем самым позволяет восстановить положение фазового перехода.

Каждая решётка рассматривается как граф ближайших соседей с фиксированными обменными интегралами $J_{ij}=1$, а конфигурация спинов задаётся набором
значений $S_i=\pm1$ на его вершинах. Таким образом, выполняется поиск отображения 
\begin{gather} \label{mapping}
\{S_i(T)\}\;\longmapsto\; y(T)\in\{\text{low},\text{high}\},
\end{gather} 
где метка $y$ кодирует принадлежность конфигурации к низкотемпературной или высокотемпературной фазе.

\begin{figure}[H]

\centering
\includegraphics[width=0.65\textwidth]{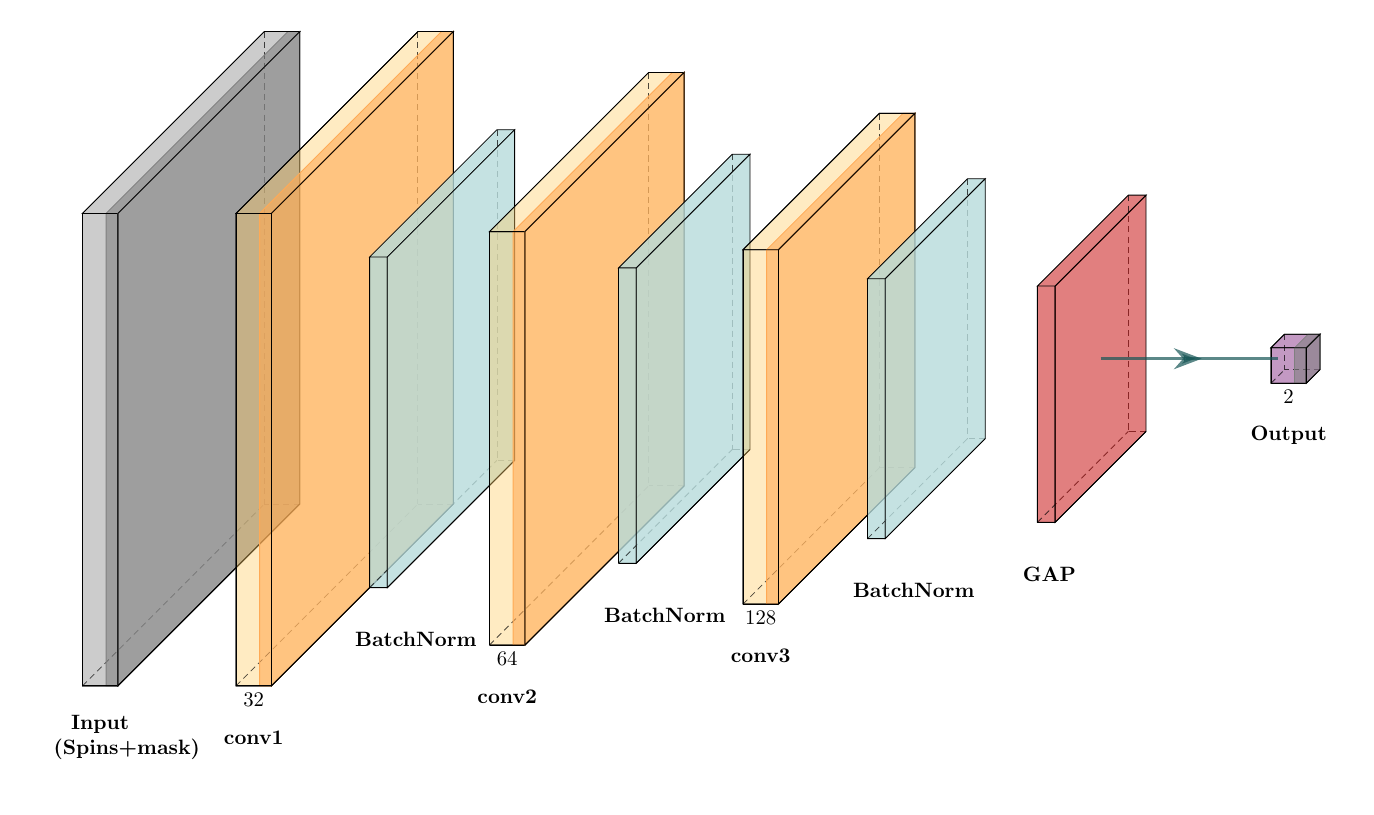}
\smallskip
\caption{Предложенная архитектура CNN3.}
\label{img:cnn3_architecture}
\end{figure}

Для подачи на вход свёрточной сети разнородные геометрии приводились к единому
формату: каждой конфигурации сопоставлялась фиксированная матрица размера
$56\times 66$ с двумя каналами. Первый канал содержал значения спинов в
кодировке $\pm 1$, второй — бинарную маску занятых позиций (1 на узлах решётки
и 0 на пустом фоне). Такой формат обеспечивает устойчивость к отличиям в
линейных размерах и позволяет одной архитектуре обрабатывать различные типы
решёток. Для квадратной и треугольной геометрий исходные матрицы спинов
центрировались в общей матрице без деформаций.

Для гексагональной решётки были реализованы два способа укладки данных.
В первом случае спины записывались построчно во множество строк общей матрицы
с последующим маскированием незаполненной области. Во втором варианте
использовалась укладка по обходу в ширину (Breadth-First Search, BFS):
из параметров элементарной ячейки восстанавливался граф ближайших соседей,
строилась перестановка узлов по обходу в ширину от фиксированного узла,
после чего конфигурация записывалась построчно с чередованием направления по
строкам. Такая перестановка уменьшает разрыв локальной связности на входе и
существенно снижает размывание кривых вероятности высокотемпературной фазы $P_{\text{high}}(T)$ вблизи $T_c$.
Для решётки кагоме аналогично применялись BFS-укладка и запись в общую матрицу
с маской. Для повышения устойчивости и учёта симметрий системы использовались
аугментации: случайные повороты конфигураций на $90^\circ$ и отражения
относительно вертикальной и горизонтальной осей, применяемые синхронно к обоим
каналам. Это формирует фактическую инвариантность модели к
основным симметриям решётки при сохранении корректной бинарной маски.

Архитектура классификатора, изображённая на рис.~\ref{img:cnn3_architecture},
представляет собой последовательность свёрточных блоков с последующей агрегацией
через слой глобального усреднения и выходным полносвязным слоем размерности два,
который возвращает вероятности низкотемпературной и высокотемпературной фаз.
Структуру сети можно записать в виде:

$Conv2D: (32, 3 \times 3)$ $\rightarrow$ $BatchNorm$ $\rightarrow$ $Dropout: (0.25)$
$\rightarrow$ $Conv2D: (64, 3 \times 3)$ $\rightarrow$ $BatchNorm$ $\rightarrow$ $Dropout: (0.25)$
$\rightarrow$ $Conv2D: (128, 3 \times 3)$ $\rightarrow$ $BatchNorm$ $\rightarrow$ $Dropout: (0.25)$
$\rightarrow$ $GAP$ $\rightarrow$ $Dense: (2)$.

Слои BatchNorm включены после свёрточных слоёв для нормализации активаций
в каждом мини-батче, что стабилизирует распределения признаков, ускоряет
сходимость обучения и выступает дополнительной регуляризацией. Оптимизация осуществлялась
методом Adam со скоростью обучения $10^{-3}$ и размером пакета 256.
Одна и та же CNN обучалась совместно на квадратной, треугольной и гексагональной
решётках. Конфигурации на решётке кагоме использовались только на этапе тестирования
обобщающей способности модели.
\begin{figure*}[t]
  \centering
  \includegraphics[width=0.49\textwidth]{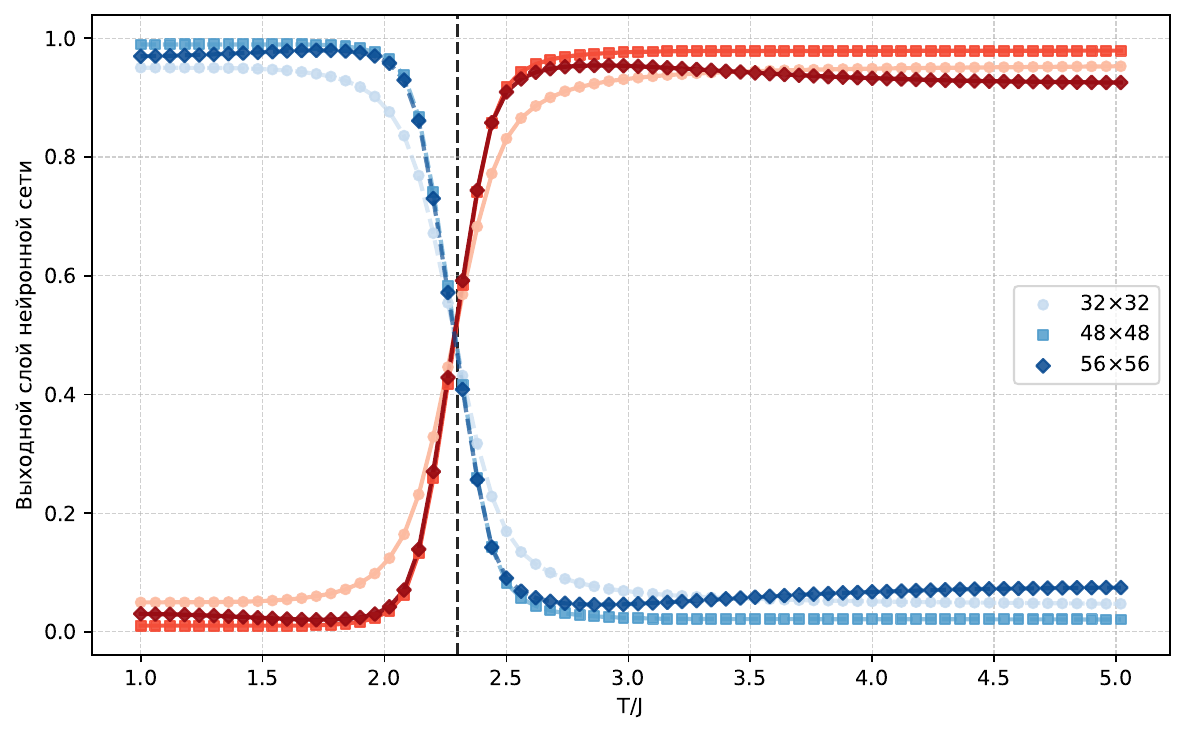}\hfill
  \includegraphics[width=0.49\textwidth]{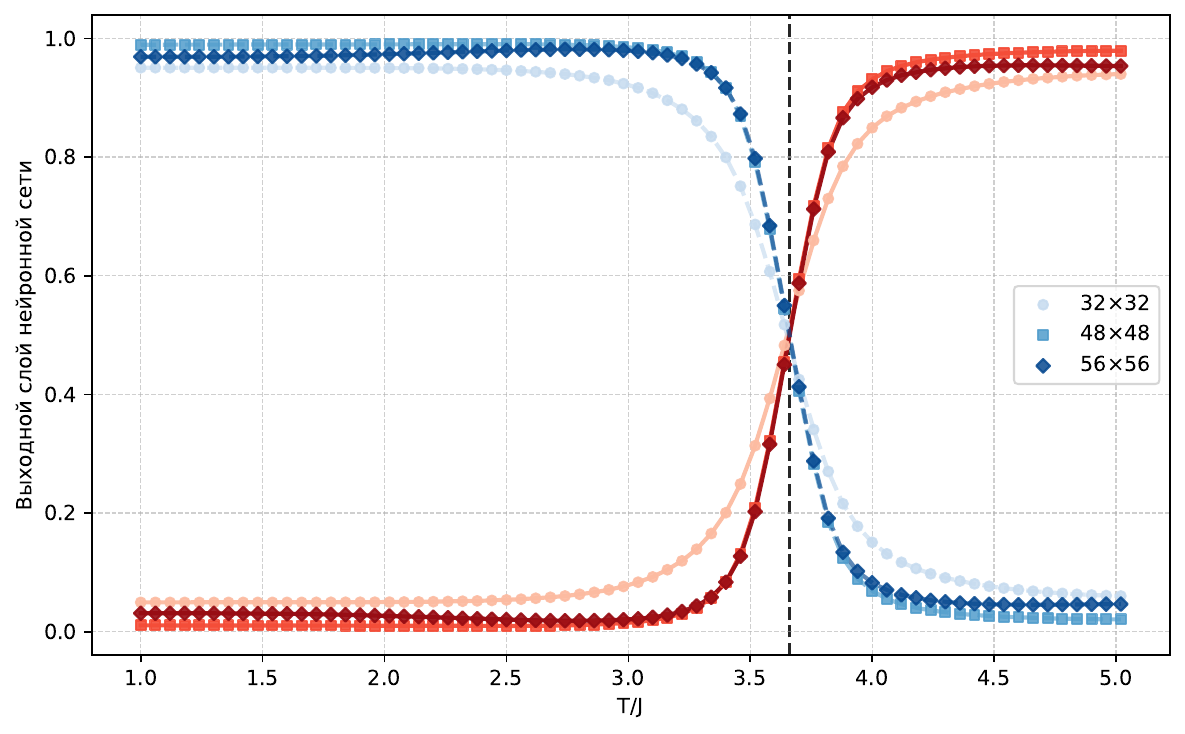}
  \caption{Квадратная и треугольная решётки. Усреднённые по конфигурациям кривые \(P_{\mathrm{high}}(T)\) (красные) и \(1-P_{\mathrm{high}}(T)\) (синие) для разных размеров. Чёрная вертикальная линия \(T_c^{\text{ref}}\).}
  \label{fig:sq-tri}
\end{figure*}

\begin{figure*}[t]
  \centering
  \includegraphics[width=0.49\textwidth]{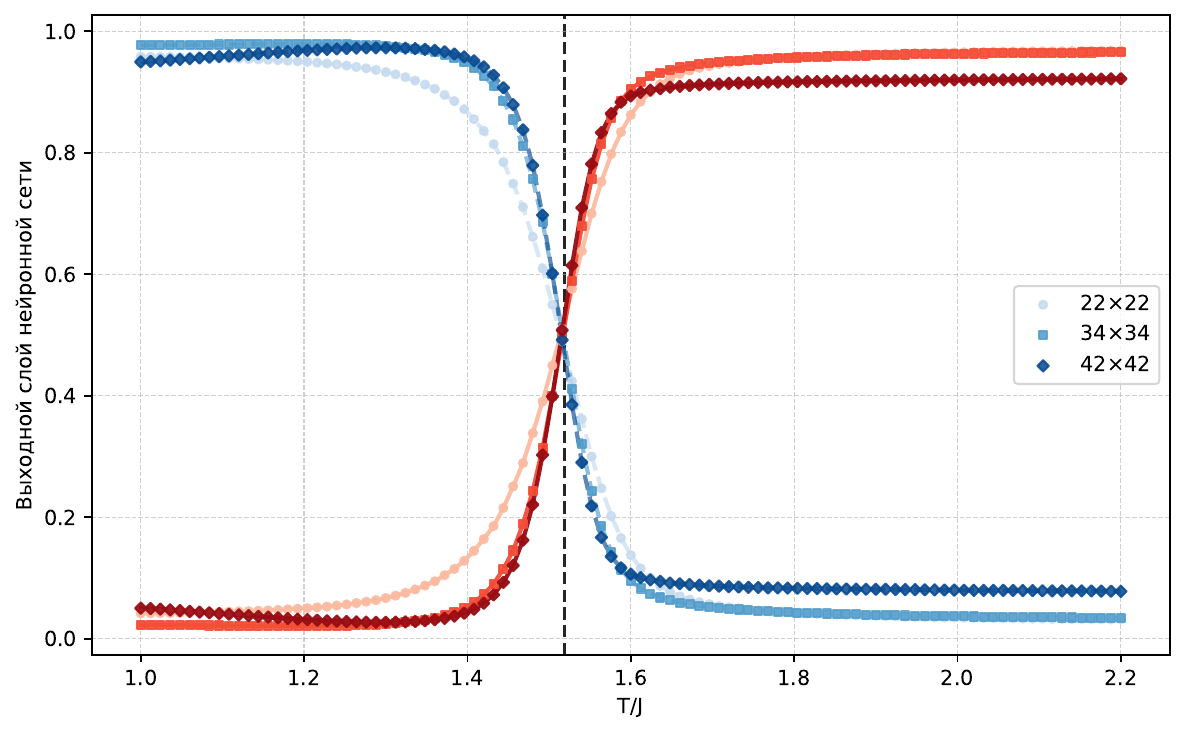}\hfill
  \includegraphics[width=0.49\textwidth]{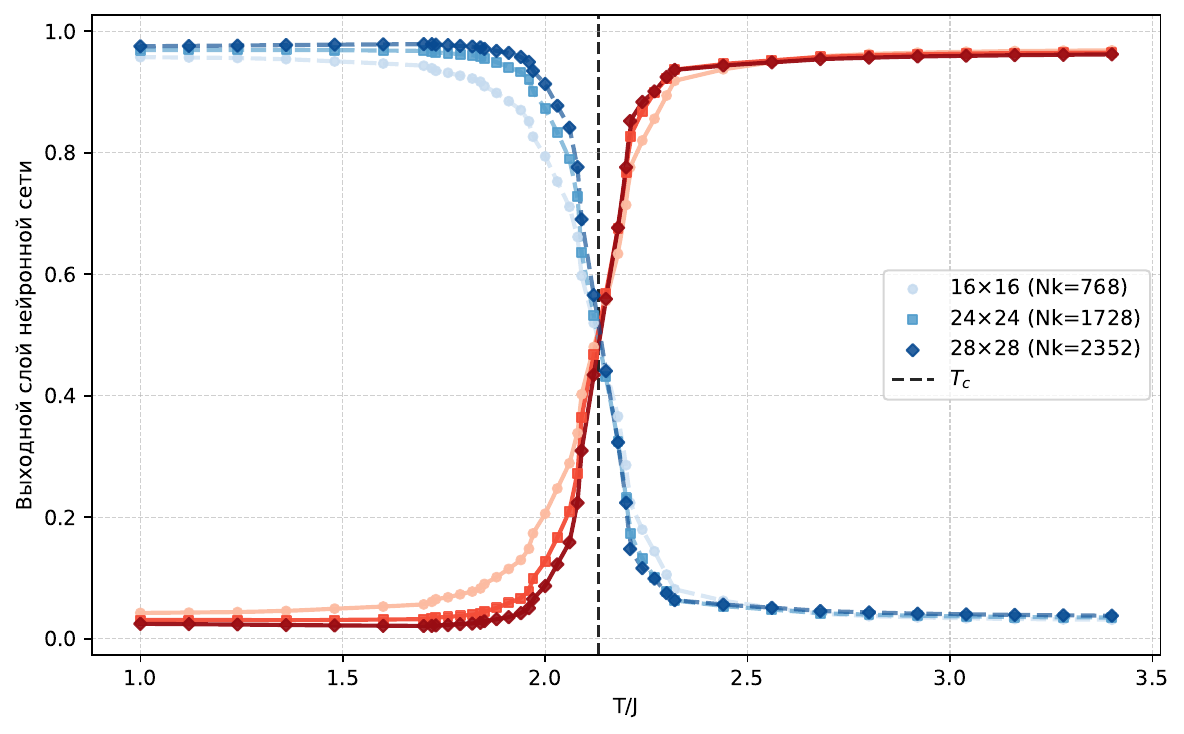}
  \caption{Гексагональная и кагоме решётки. Усреднённые по конфигурациям кривые \(P_{\mathrm{high}}(T)\) (красные) и \(1-P_{\mathrm{high}}(T)\) (синие) для разных размеров. Чёрная вертикальная линия \(T_c^{\text{ref}}\).}
  \label{fig:hex-kag}
\end{figure*}

\begin{figure}[t]
  \centering
  \includegraphics[width=0.70\linewidth]{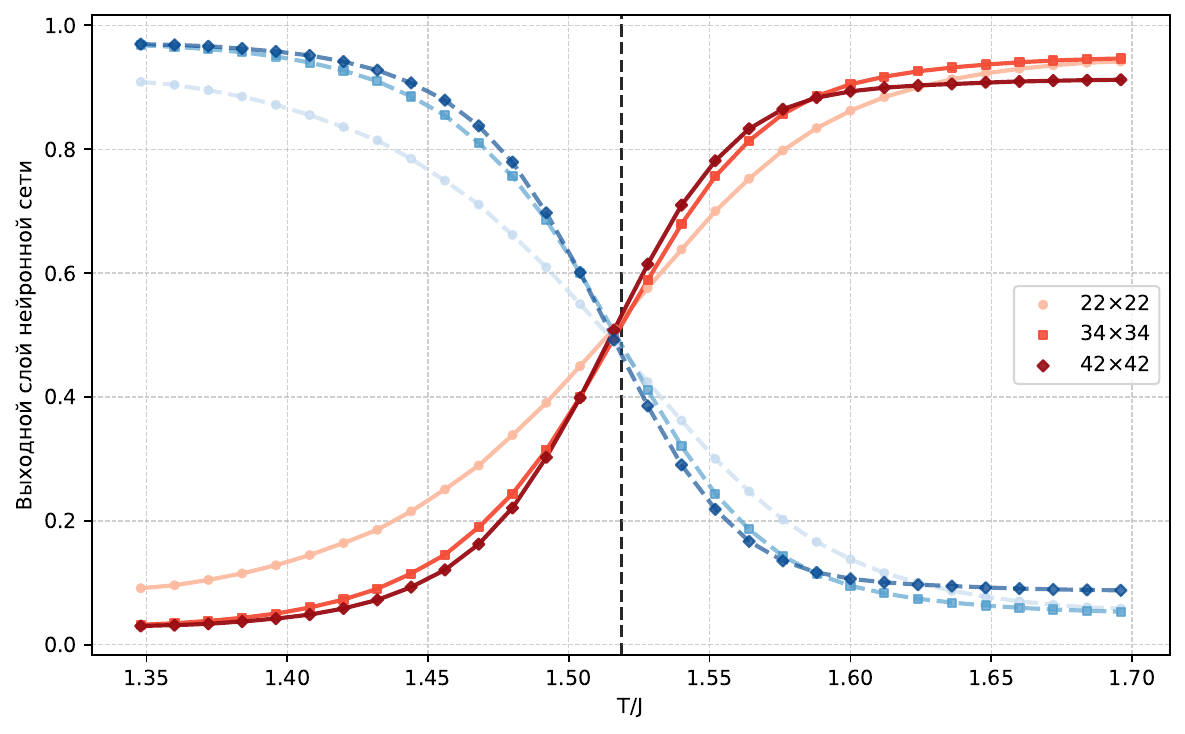}
  \caption{Предсказание нейронной сети для гексагональной решётки увеличенная окрестность \(T_c\). Чёрная вертикальная линия \(T_c^{\text{ref}}\).}
  \label{fig:hex-zoom}
\end{figure}

Значением на выходе нейронной сети является вероятность высокотемпературной фазы,
которую обозначим через $P_{\text{high}}(x)$ для отдельной конфигурации $x$.
Для фиксированной температуры $T$ рассматривается усреднённая по конфигурациям
зависимость
\begin{gather}
P_{\text{high}}(T)=\frac{1}{n_T}\sum_{i=1}^{n_T}
\bigl[\mathrm{CNN}(x_i(T))\bigr]_{\text{класс}=\text{high}},
\end{gather}
где усреднение проводится по $n_T$ независимым конфигурациям $x_i(T)$,
сгенерированным при данной температуре. Функция $P_{\text{high}}(T)$ играет роль
эффективного «порядкового параметра» классификатора: при низких температурах
она близка к нулю, а при высоких — к единице. В качестве оценки критической
температуры используется решение уравнения
\begin{gather}
P_{\text{high}}(T)=\frac{1}{2},
\end{gather}
полученное с помощью локальной линейной интерполяции по $T$. Для анализа
сопоставлялись также кривые $1-P_{\text{high}}(T)$; пересечение
$P_{\text{high}}(T)$ и $1-P_{\text{high}}(T)$ вблизи $T_c$ наглядно иллюстрирует смену преобладающей фазы. На рис.~\ref{fig:sq-tri} и~\ref{fig:hex-kag} показаны усреднённые кривые $P_{\text{high}}(T)$ и $1-P_{\text{high}}(T)$ для различных размеров и геометрий решёток, а на рис.~\ref{fig:hex-zoom} приведена увеличенная окрестность критической области для гексагональной решётки.

Для независимой проверки результатов классификатора вычислялась теплоёмкость
на спин:
\begin{gather}
C(T)=N\,\frac{\langle E^2\rangle_T-\langle E\rangle_T^2}{T^2},
\end{gather}
где $N$ — число спинов, а $\langle E\rangle_T$ и $\langle E^2\rangle_T$ —
соответственно первая и вторая моменты энергии при температуре $T$ \cite{fisher1988,jesi2016}.
\begin{figure}[h]
    \centering
    \subfloat[Квадратная решётка]{
        \includegraphics[width=0.45\textwidth]{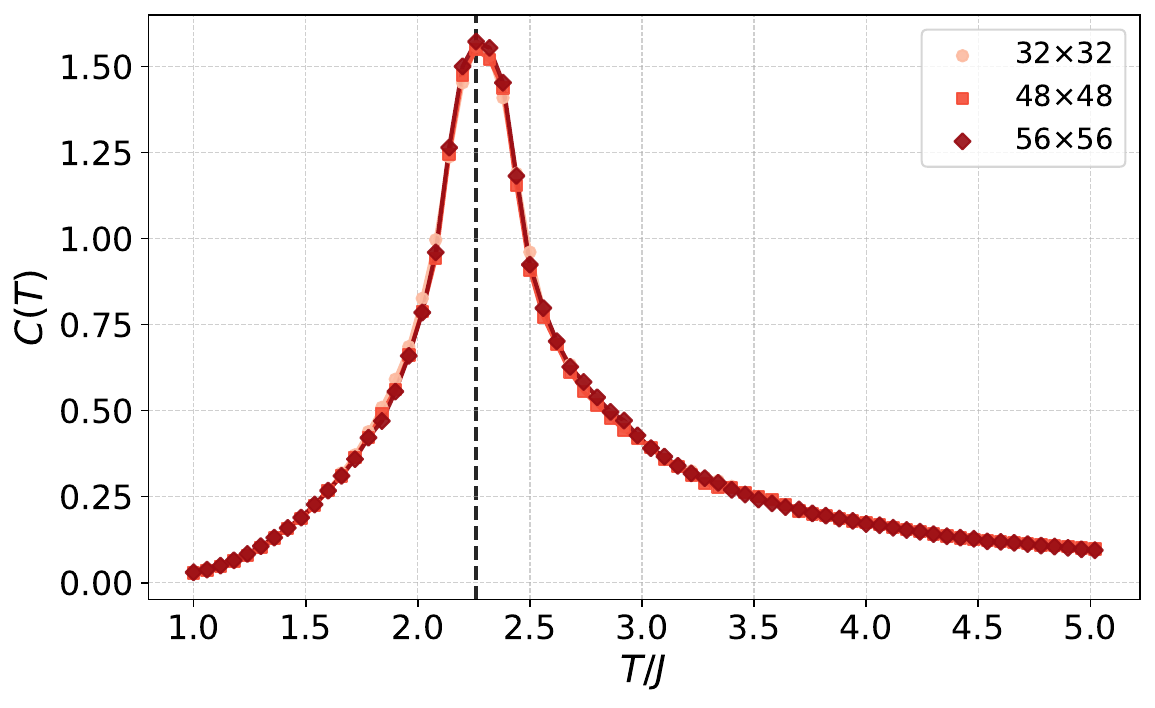}
        \label{fig:Cv_square}
    }\hfill
    \subfloat[Треугольная решётка]{
        \includegraphics[width=0.45\textwidth]{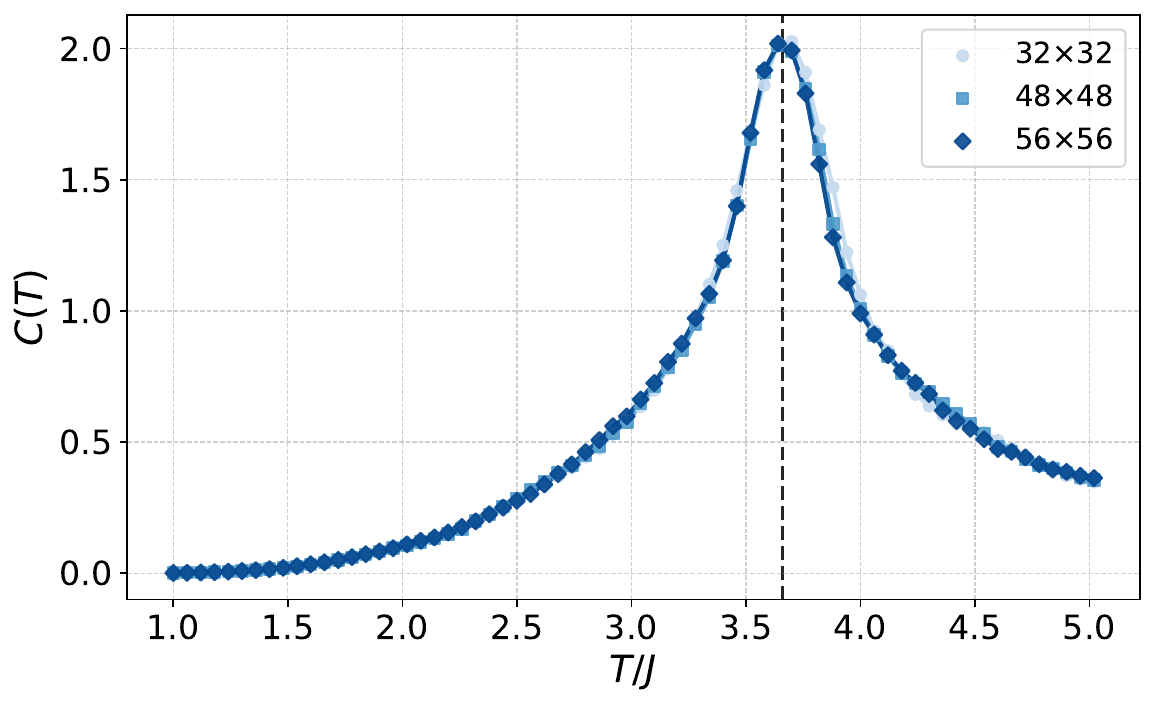}
        \label{fig:Cv_triangular}
    }\hfill
    \subfloat[Гексагональная решётка]{
        \includegraphics[width=0.45\textwidth]{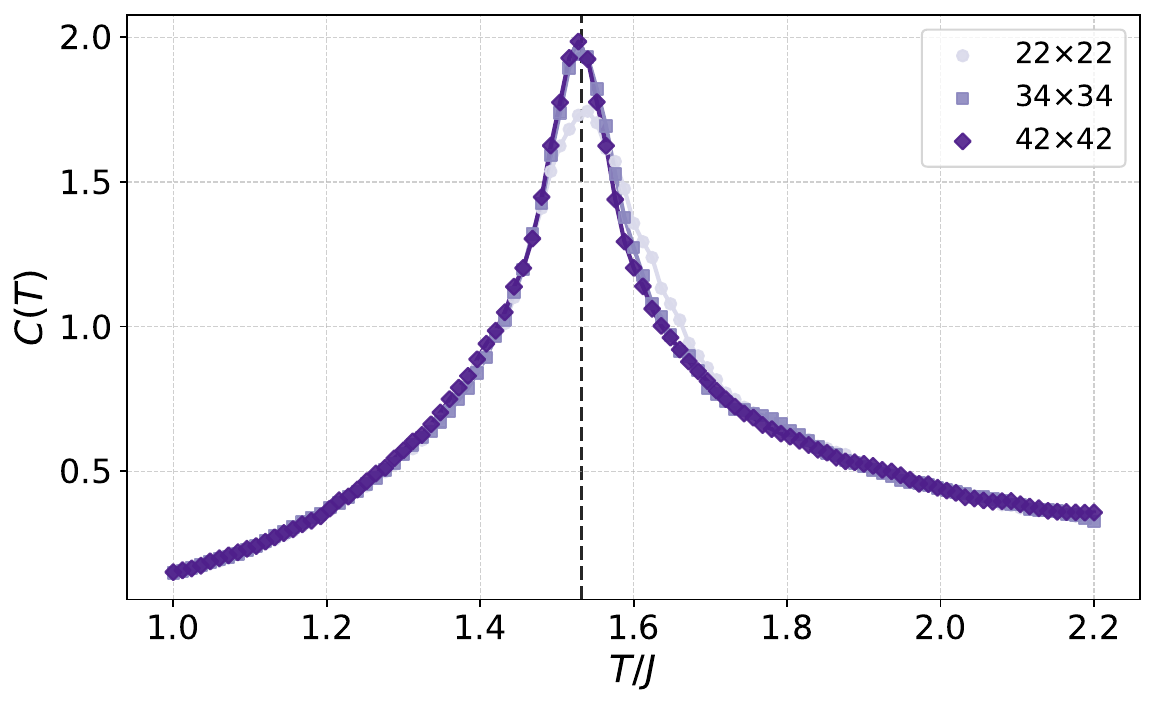}
        \label{fig:Cv_honeycomb}
    }
    \caption{$C(T)$ в зависимости от $T/J$ для (a) квадратной, (b) треугольной и (c) гексагональной решёток при разных размерах системы. Чёрная прерывистая линия обозначает максимум $C(T)$.}
    \label{fig:Cv_all_lattices}
\end{figure}
Максимум $C(T)$ сопоставлялся с положением пересечения $P_{\text{high}}(T)$ (рис.~\ref{fig:Cv_all_lattices}).
В качестве контрольных 
значений \cite{onsager1944,houtappel1950,syozi1951,baxter1982} критической температуры ферромагнитной модели Изинга использовались
\begin{gather}
T_c^{\text{sq}}\approx 2.269,\quad
T_c^{\text{tri}}\approx 3.641,\quad
T_c^{\text{hc}}\approx 1.519,\quad
T_c^{\text{kag}}\approx 2.143.
\end{gather}
В таблице~\ref{tab:deltas} приведены отклонения предсказанных нейронной сетью значений $T_c$ 
от $T_c^{\text{ref}}$ для различных геометрий и размеров решёток.
\begin{table*}[h]
  \centering
  \caption{Отклонения предсказанных значений $T_c$ нейронной сетью от \(T_c^{\text{ref}}\) для разных решёток и размеров.}
  \label{tab:deltas}
  \begin{tabularx}{\textwidth}{|l|*{3}{>{\centering\arraybackslash}X|}}
    \hline
    Геометрия решётки & \(32\times 32\) & \(48\times 48\) & \(56\times 56\) \\
    \hline
    Квадратная      & \(+0.006\) & \(-0.010\) & \(+0.006\) \\
    \hline
    Треугольная     & \(-0.009\) & \(-0.001\) & \(+0.002\) \\
    \hline
    Гексагональная  & \(-0.005\) & \(-0.002\) & \(-0.004\) \\
    \hline
    Кагоме          & \(-0.007\) & \(+0.001\) & \(+0.006\) \\
    \hline
  \end{tabularx}
\end{table*}

Полученные результаты показывают, что на квадратной, треугольной и гексагональной
решётках, CNN формирует отчётливые S-образные кривые $P_{\text{high}}(T)$
с пересечением вблизи соответствующих критических температур. Переход от
простого построчного представления к BFS-укладке для гексагональной решётки
устраняет систематический сдвиг по температуре и повышает крутизну кривых в
критической области, что согласуется с положением максимумов $C(T)$.
При применении обученного классификатора к решётке кагоме пересечение
$P_{\text{high}}(T)$ локализуется в окрестности
\begin{gather}
T_c^{\text{kag}}=\frac{4}{\ln(3+2\sqrt{3})}\approx 2{.}143,
\end{gather}
что подтверждает переносимость выученных признаков между различными геометриями
при геометрически информированной укладке данных на вход. Сопоставление
построчного представления и BFS-перестановки показывает, что геометрически
осмысленная укладка является ключевым фактором снижения размывания решения
в критической области и уменьшения систематического сдвига оценки $T_c$.
Используемая регуляризация стабилизирует обучение при совместной настройке
на нескольких геометриях и способствует унификации извлекаемых признаков.

Таким образом, единый свёрточный классификатор позволяет надёжно выявлять
фазовые переходы в двумерных решётках различной геометрии и без дополнительного
обучения переносится на решётку кагоме. Оценки критических температур,
извлечённые из пересечения $P_{\text{high}}(T)$, хорошо
согласуются как с теоретическими значениями $T_c^{\text{ref}}$, так и с
положением максимумов теплоёмкости $C(T)$. Канал маски и геометрически
информированная укладка позволяют уменьшить размывание
в критической области без усложнения архитектуры модели.

\paragraph{Заключение }

Проведённое исследование демонстрирует эффективность применения свёрточных нейронных сетей к изучению низкотемпературных фаз спиновых систем и фазовых переходов в фрустрированных магнитных материалах. Предложенные подходы на основе CNN позволили максимально  эффективно решить две задачи: восстановление зависимости средней энергии от пространственного распределения обменных интегралов и универсальную классификацию фазовых состояний без требования явного вычисления статистической суммы.

Результаты свидетельствуют о том что CNN-архитектуры  превосходят традиционные полносвязные и глубокие нейронные сети с кастомной архитектурой  в точности предсказаний, особенно при низких температурах, где достигается высокая точность при $T<1$. Разделение обменных интегралов на каналы горизонтальных $J_{hor}$ и вертикальных $J_{ver}$ взаимодействий обеспечило эффективный учёт пространственной структуры системы, что является критическим для выявления сложных связей между термодинамическими характеристиками и структурой  в низкотемпературной фазе.

Важным результатом является универсальность разработанного классификатора, способного различать низко- и высокотемпературные состояния на решётках различной геометрии (квадратная, треугольная, гексагональная, кагоме) без переобучения. Использование единой фиксированной матрицы размером $56\times66$ с бинарной маской занятых позиций обеспечило устойчивость алгоритма к различиям в размерах и геометриях. Применение кластерного алгоритма Свендсена-Ванга для генерации независимых конфигураций и исключение данных из области критической температуры предотвратили переобучение и обеспечили надёжность валидации.

Предложенные подходы открывают возможности для расширения исследований на более широкий класс спиновых систем, включая модели с иными архитектурами, типами взаимодействия и размерностями, а также для анализа других термодинамических характеристик. Решения на основе сверточных нейронных сетей демонстрируют отличный потенциал методов машинного обучения в решении классических проблем статистической физики, особенно в ситуациях, когда экспоненциальный рост числа микросостояний делает прямые вычисления практически недостижимыми. Дальнейшие исследования могут быть направлены на применение более сложных архитектур нейронных сетей, включая графовые нейронные сети для явного учёта топологии взаимодействий, а также на расширение подхода на трёхмерные системы и анизотропные взаимодействия.


\label{sect:bib}

\bibliographystyle{plain}





\end{document}